\documentclass[aps,prb,twocolumn,superscriptaddress,bibliography]{revtex4-2}
\usepackage{amsfonts}
\usepackage{mathrsfs}
\usepackage{amsmath}
\usepackage{color}
\usepackage{graphicx}
\usepackage{bm}
\usepackage{amssymb}
\usepackage{xspace}
\usepackage{epstopdf}
\usepackage{dcolumn}
\usepackage{multirow}
\usepackage{float}
\usepackage{comment}

\usepackage[colorlinks=true, letterpaper=true, pdfstartview=FitV,  linkcolor=blue, citecolor=blue, urlcolor=blue]{hyperref}

\makeatother

\begin{document}

\title{Ideal spin-polarized Weyl-half-semimetal with a single pair of Weyl points in half-Heusler compounds $X$CrTe ($X$=K, Rb)}

\author{Hongshuang Liu}
\thanks{These authors contributed equally to this work}
\affiliation{Tianjin Key Laboratory of Low Dimensional Materials Physics and Preparation Technology, School of Science, Tianjin University, Tianjin 300354, China}

\author{Jin Cao}
\thanks{These authors contributed equally to this work}
\affiliation{Research Laboratory for Quantum Materials, IAPME, FST, University of Macau, Macau SAR, China}

\author{Zeying Zhang}
\affiliation{College of Mathematics and Physics, Beijing University of Chemical Technology, Beijing 100029, China}

\author{Jiashuo Liang}
\affiliation{Tianjin Key Laboratory of Low Dimensional Materials Physics and Preparation Technology, School of Science, Tianjin University, Tianjin 300354, China}

\author{Liying Wang}
\email{liying.wang@tju.edu.cn}
\affiliation{Tianjin Key Laboratory of Low Dimensional Materials Physics and Preparation Technology, School of Science, Tianjin University, Tianjin 300354, China}
\affiliation{Tianjin Demonstration Center for Experimental Physics Education, School of Science, Tianjin University, Tianjin 300354, China}

\author{Shengyuan A. Yang}
\email{yangshengyuan@um.edu.mo}
\affiliation{Research Laboratory for Quantum Materials, IAPME, FST, University of Macau, Macau SAR, China}


\begin{abstract}
Realizing ideal Weyl semimetal state with a single pair of Weyl points has been a long-sought goal in the field of topological semimetals. Here, we reveal such a state in the Cr-based half-Heusler compounds $X$CrTe ($X$=K, Rb). We show that these materials have a half metal ground state, with Fermi level crossing only one spin channel. Importantly, the Fermi surface is clean, consisting of the minimal number (i.e., a single pair) of spin-polarized Weyl points, so the state represents an ideal Weyl half semimetal. We show that the locations of the two Weyl points and the associated Chern vector can be flexibly tuned by rotating the magnetization vector. The minimal surface Fermi arc pattern and its contribution to anomalous Hall transport are discussed. Our finding offers an ideal material platform for exploring magnetic Weyl fermions, which will also facilitate the interplay between Weyl physics and spintronics. 
\end{abstract}
\maketitle

\section{Introduction}
The study of topological semimetals and novel emergent quasiparticles in materials have been a focus of research in the past decade~\cite{Chiu2016Classification,Yan2017Topological,Armitage2018Weyl,Lv2021Experimental}. Weyl semimetals are one most prominent class of topological semimetals~\cite{Armitage2018Weyl,Wan2011Topological,Murakami2007Phase}. In Weyl semimetals, the low-energy bands cross at isolated twofold degenerate nodal points, known as Weyl points. Electron quasiparticles around these points behave as Weyl fermions, and the Weyl points act like monopoles in momentum space, emitting or absorbing Berry curvature field. The existence of Weyl points requires the breaking of either time-reversal symmetry $T$ or inversion symmetry $P$, and the no-go theorem dictates that Weyl points must occur in pairs with opposite chirality~\cite{Nielsen1981Absence,Nielsen1981Absencea}. These symmetry and topology conditions impose constraints on the minimal number of Weyl points for a Weyl semimetal state, namely, for magnetic (nonmagnetic) systems, the allowed number of Weyl points is at least two (four)~\cite{Armitage2018Weyl}.

A long-sought goal in the research on topological semimetals is to find concrete materials realizing the simplest Weyl semimetal state — the state with the minimal number, i.e., a single pair, of Weyl points. This is desired as a minimal system for studying Weyl fermion physics. Besides the requirement on the minimal number of Weyl points, to qualify as an ideal topological semimetal material in general, the candidate should also satisfy the following conditions. First, the Weyl semimetal state should occur at the intrinsic ground state of the candidate material, without need of external fields or extrinsic tuning. Second, the energy of Weyl points should be close to the Fermi level, better sitting exactly at Fermi level without doping. Third, the Fermi surface should be clean, meaning that there are no other extraneous bands crossing the Fermi level. These conditions ensure the dominance of Weyl fermion contribution in various material properties, which facilitates the exploration of signatures of Weyl semimetals and the interpretation of experimental result. 

As mentioned above, Weyl semimetals with a single pair of Weyl points can only occur in magnetic materials. To date, only a few candidates were proposed, such as HgCr$_2$Se$_4$~\cite{Xu2011Chern}, MnBi$_2$Te$_4$~\cite{Zhang2019Topological,Li2019Intrinsic}, EuCd$_2$As$_2$~\cite{Wang2019Single,Soh2019Ideal,Roychowdhury2023Anomalous}, K$_2$Mn$_3$(AsO$_4$)$_3$~\cite{Nie2022Magnetic}, and MnSn$_2$Sb$_2$Te$_6$~\cite{Gao2023Intrinsic}. However, these examples are not perfect. They more or less have certain drawbacks. For example, the Fermi surface in HgCr$_2$Se$_4$ and K$_2$Mn$_3$(AsO$_4$)$_3$ is not clean, having other bands coexisting with Weyl points. MnBi$_2$Te$_4$ and EuCd$_2$As$_2$ require external field, doping, or strain tuning to be turned into Weyl semimetal state. The Weyl points in MnSn$_2$Sb$_2$Te$_6$ are quite close to each other and are not exactly at Fermi level. Thus, it remains an important task to search for suitable material platforms that can realize the ideal Weyl semimetal phase. 

\begin{table*}
\caption{\label{tab_i}Calculation results for of KCrTe and RbCrTe. These include the type of half-Heusler configuration with lowest energy, optimized lattice constants $a$~(\AA), cohesive energies $E_{\mathrm{coh}}$~(eV), energy difference between FM and AFM configurations $\Delta E_{\mathrm{FM-AFM}}$ (meV$/$f.u.), energy difference between magnetization vector along $[001]$ and $[110]$ directions in FM state $\Delta E_{\mathrm{100-110}}$ ($\mu$eV$/$f.u.), energy difference between magnetization vector along $[111]$ and $[110]$ directions in FM state $\Delta E_{\mathrm{111-110}}$  ($\mu$eV$/$f.u.), magnetic moment at Cr site $M$ ($\mu_{B}$/Cr), and Curie temperature $T_{\text{C}}$~(K).}
\begin{ruledtabular}
\begin{tabular}{ccccccccc}
  Systems & Type & $a$ & $E_{\mathrm{coh}}$ & $\Delta E_{\mathrm{FM-AFM}}$ & $\Delta E_{\mathrm{100-110}}$ & $\Delta E_{\mathrm{111-110}}$ & $M$ & $T_{\text{C}}$ \\
\hline 
 KCrTe & Type-I & 7.16 & -8.10 & -42.50 & -2.83 & -4.06 & 4.4 & 202 \\
RbCrTe & Type-I & 7.36 & -7.75 & -47.26 &  ~0.09 &  ~8.16 & 4.4 & 217 \\
\end{tabular}
\end{ruledtabular}
\end{table*}

In this work, based on first principles calculations and theoretical analysis, we reveal ideal Weyl half-semimetal state in half-Heusler compounds $X$CrTe ($X$=K, Rb). We show that these materials have ferromagnetic ground state with Curie temperature above 200~K. They are half semimetals, meaning that the low-energy states belong to a single spin channel, whereas the other spin channel is gapped. A pair of Weyl points are located exactly at Fermi level enforced by band filling, and the Fermi surface is clean without other extraneous bands. The two Weyl points are well separated in momentum space, and their locations can be tuned by rotating the magnetization vector. The ideal bulk Weyl band structure also leads to the simplest topological Fermi arc pattern on sample surfaces, and its contribution to anomalous Hall transport is discussed.

\section{Computation details}
Our first-principles calculations were based on density functional theory (DFT), by using the Vienna \textit{ab initio} Simulation Package (VASP)~\cite{Bloechl1994Projector,Kresse1999ultrasoft,Kresse1996Efficient,Kresse1996Efficiency}. The exchange-correlation potential was treated within the generalized gradient approximation (GGA) with the Perdew–Burke–Ernzerhof (PBE) implementation~\cite{Perdew1996Generalized}. The cut-off energy was set as 500~eV. The Brillouin zone (BZ) was sampled with the $\Gamma$-centered $k$ mesh of size $15\times15\times15$. The force and energy convergence criteria were set as $0.01$~eV/\AA~and 10$^{-6}$~eV, respectively. The possible correlation effects of Cr-$3d$ orbitals were treated by the DFT$+$U method~\cite{Dudarev1998Electron}. We tested the $U$ values from 0 to 5~eV and found no qualitative difference regarding the low-energy bands (see Supplementary Materials). Hence, in the following, we focus on the $U=0$ results. The phonon dispersion was calculated by using PHONOPY code~\cite{Togo2015First}. Thermal stability was investigated by performing finite-temperature \textit{ab initio} molecular dynamics (AIMD) simulations~\cite{Pastore1991Theory}. The Curie temperature was estimated by Monte Carlo simulations using the VAMPIRE software package~\cite{Evans2014Atomistic}. To investigate surface states, we constructed maximally localized Wannier functions by using the Wannier90 code~\cite{Marzari1997Maximally,Souza2001Maximally,Mostofi2008wannier90}, and the surface Green functions were computed by the iterative algorithm~\cite{Sancho1984Quick,Sancho1985Highly}, as implemented in the WannierTools package~\cite{Wu2018WannierTools}.

\section{Results and discussion}

\begin{figure}[b]
\begin{centering}
\includegraphics[width=8.6cm]{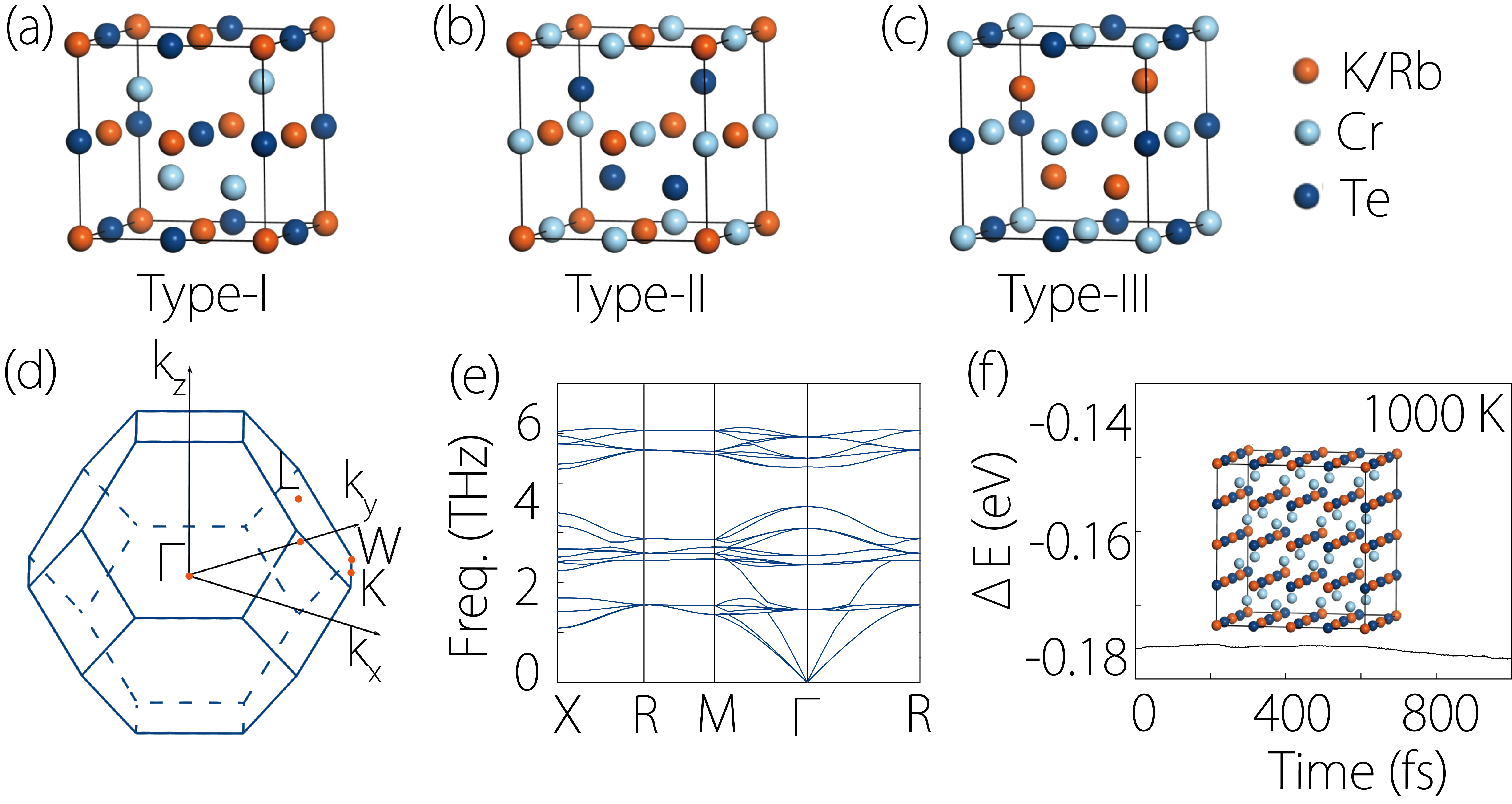}
\par\end{centering}
\caption{\label{Fig_1}Lattice structures of half-Heusler compounds KCrTe and RbCrTe in (a) type-I, (b) type-II, and (c) type-III configurations. (d) Brillouin zone with high-symmetry points labeled. (e) Calculated phonon spectrum of KCrTe. (f) Total energy ($E_t$) variation of KCrTe versus time in AIMD simulations at 1000~K.$\Delta$E = $E_{t}$ + 393.1 eV. The inset shows the snapshot of lattice structure at the end of simulation period.}
\end{figure}

\subsection{Crystal structure and stability}
Half Heusler compounds are a huge family of materials. They are intermetallics with face-centered cubic crystal structure. Their composition is of the form $XYZ$, where $X$ and $Y$ are metal elements and $Z$ is a p-block element. Half Heusler materials have diverse properties. Many of them exhibit excellent properties desired for spintronics, such as various magnetic orderings, large magneto-resistance, half metallicity, and etc~\cite{Heusler1903Uber,Manna2018Heusler,Yan2014Half}. Previous studies also revealed several Heusler materials as topological insulators~\cite{Xiao2010Half,Liu2016Observation}.

Here, we focus on the half Heusler compounds $X$CrTe ($X$=K, Rb). For half Heusler compounds, there are three possible types of atomic configurations, arising from the distinct atomic arrangement in the lattice, as shown in Fig.~\ref{Fig_1}(a-c). (They all share the $C_{1b}$ structure with space group of $F\bar{4}3m$ (No.~216).) In the three configurations, Cr atoms are surrounded (a) by four $X$ and four $Z$ atoms (type-I), (b) by four $Z$ atoms (type-II), and (c) by four $X$ atoms (type-III). From our DFT calculation, we find that the most stable configuration is type-I for KCrTe and RbCrTe (see Fig.~S1 in the Supplementary Materials). The optimized lattice constants of KCrTe and RbCrTe are 7.16~\AA~and 7.36~\AA, respectively. The relevant information is also shown in Table~\ref{tab_i}. Because the results of KCrTe and RbCrTe are qualitatively similar, in the main text, we will mainly focus on KCrTe. The corresponding results of RbCrTe are presented in the Supplementary Materials.

The structural stability of $X$CrTe ($X$=K, Rb) is investigated by computing the cohesive energy and phonon spectra, and by AIMD simulations. The cohesive energy ($E_{\mathrm{coh}}$) is calculated as $E_{\mathrm{coh}}=E_{\mathrm{KCrTe/RbCrTe}}-E_{\mathrm{K/Rb}}-E_{\mathrm{Cr}}-E_{\mathrm{Te}}$, where $E_{\mathrm{KCrTe/RbCrTe}}$ is the total energy per formula unit (f.u.) for KCrTe/RbCrTe, $E_{\mathrm{K/Rb}}$, $E_{\mathrm{Cr}}$, and $E_{\mathrm{Cr}}$ are the isolated atomic energies of K/Rb, Cr, and Te, respectively. The obtained negative cohesive energy listed in Table~\ref{tab_i} is quite large ($>7$~eV/f.u.), suggesting good energetic stability of $X$CrTe ($X$=K, Rb). The calculated phonon spectrum for KCrTe is shown in Fig.~\ref{Fig_1}(e), with no imaginary frequency throughout the entire BZ. This confirms the dynamical stability of the material. Furthermore, the thermal stability of the material is verified by the AIMD simulations. We perform the simulation for temperatures up to 1000~K and find that these compounds still maintain their structural integrity (see Fig.~\ref{Fig_1}(f)), confirming their excellent thermal stability.

\begin{figure}
\begin{centering}
\includegraphics[width=8.6cm]{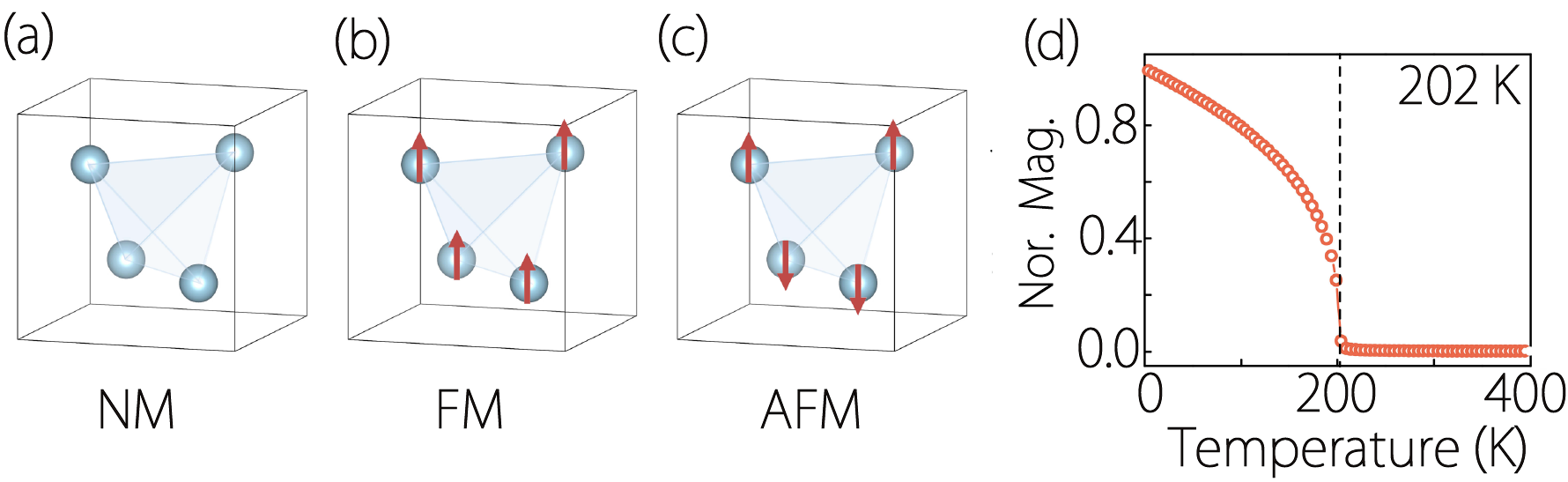}
\par\end{centering}
\caption{\label{Fig_2}Schematics of (a) nonmagnetic, (b) ferromagnetic, and (c) antiferromagnetic configurations of $X$CrTe ($X$=K, Rb). (d) Monte Carlo simulation result for the temperature dependence of magnetization of KCrTe.}
\end{figure}

\subsection{Magnetic configuration}
Half Heusler compounds often exhibit magnetic ordering. The materials $X$CrTe ($X$=K, Rb) studied here contain transition metal elements Cr, which possesses magnetic moment due to its partially filled 3$d$ shell. We determine their magnetic ground state by comparing the total energies of three typical magnetic configurations for such half Heusler structure, including paramagnetic (NM), ferromagnetic (FM), and antiferromagnetic (AFM) configurations, as illustrated in Fig.~\ref{Fig_2}(a-c). By comparing the calculated total energies (see Table~\ref{tab_ii}), we find $X$CrTe ($X$=K, Rb) both prefer the FM ground state, with relatively large energy difference ($>40$~meV/f.u.) between the FM and AFM configurations. In the FM state, we find that magnetic moments are mainly distributed on the Cr site, with a large value $\sim$4.4~$\mu_{B}$/Cr. This agrees with the nominal valence of Cr being $+1$, with five 3$d$ electrons in the same spin state.

Next, we investigate the magnetic anisotropy of the FM ground state. To this end, we calculate and compare the energies for magnetization vector along $[001]$, $[110]$, and $[111]$ high symmetry directions, with spin-orbit coupling (SOC) included. The energy differences are shown in Table~\ref{tab_i}. One observes that in these materials, the magnetic anisotropy is weak, of just a few $\mu$eV/f.u.. This is on the same order of magnitude as typical isotropic magnet like Fe, Co, and Ni~\cite{Halilov1998Magnetocrystalline,Laan1998Microscopic}. Such weak anisotropy is typical for cubic crystal systems, which have relative high symmetry. The soft magnetism in $X$CrTe ($X$=K, Rb) suggests that the magnetization direction in these materials may be readily controlled, e.g., by applied magnetic field. As we see later, this is an advantage, which allows a flexible control of the Weyl points (and the topological band structure) by rotating the magnetization vector. 

We estimate the Curie temperatures of $X$CrTe ($X$=K, Rb) by Monte Carlo simulations, based on a classical Heisenberg type spin model:
\begin{equation}
H=\sum_{ij} J_{ij} \boldsymbol{S}_i \cdot \boldsymbol{S}_j,
\end{equation}
where $\boldsymbol{S}_i$ is the unit spin vector at site $i$, and $J_{ij}$ is strength of magnetic exchange interaction between sites $i$ and $j$. In the model, we neglect the magnetic anisotropy energy term, due to its weak strength, as discussed just now. We also tested that including anisotropy terms have little influence on the estimated Curie temperature. As for the exchange terms, we include exchange parameters up to second neighbors (see Supplementary Materials for the derivation). The magnetization curve simulated for KCrTe is plotted in Fig.~\ref{Fig_2}(d). One observes that its Curie temperature is about 202~K. The Curie temperature for RbCrTe is similar $\sim$217~K. These values are much higher than liquid nitrogen temperature and also higher than that of the reported magnetic topological semimetal Co$_3$Sn$_2$S$_2$ ($\sim$175~K). 

\begin{table}
\caption{\label{tab_ii}Locations (in units of \AA$^{-1}$) and chiralities of the two Weyl points in KCrTe, when the magnetization vector is along $[001]$ direction.}
\begin{ruledtabular}
\begin{tabular}{cccccc}
  Direction & Weyl points & $k_x$ & $k_y$ & $k_z$ & Chirality \\
\hline 
$[001]$ & $W_1$ & 0.0 & 0.0 &  ~0.066 & $-1$ \\
        & $W_2$ & 0.0 & 0.0 & -0.066 & $+1$ \\
\end{tabular}
\end{ruledtabular}
\end{table}

\subsection{Electronic band structure}
Let’s first consider the electronic band structure in the absence of SOC. The result for FM KCrTe along with the projected density of states (PDOS) are shown in Fig.~\ref{Fig_3}(a). The results for RbCrTe are very similar and are given in the Supplementary Materials. In the absence of SOC, the two spin channels are decoupled. One observes that the spin-down channel is semiconducting with a large energy gap of about 3~eV. Meanwhile, the spin-up channel shows a semimetal character, where the conduction and the valence bands touch at the $\Gamma$ point on the Fermi level. Such a system is simultaneously a half metal and a semimetal, and hence is referred to as a half semimetal~\cite{Liu2015Multiple}. From the PDOS plot, one can see the low-energy bands around Fermi level are mainly from Cr-3$d$ orbitals.

For the band structure in Fig.~\ref{Fig_3}(a), the Fermi point is a band touching point at $\Gamma$. Note that this point is threefold degenerate: it is formed by one conduction band with two valence bands. The two valence bands are degenerate along the $\Gamma$-Z and $\Gamma$-L paths. The band dispersion around this point is of quadratic type. Such kind of band touching point was investigated by Zhu \textit{et al.}~\cite{Zhu2018Quadratic} before and was termed as the quadratic contact point. In the present system, the quadratic contact point is protected by the $T_d$ point group symmetry at $\Gamma$.

\begin{figure}
\begin{centering}
\includegraphics[width=6.6cm]{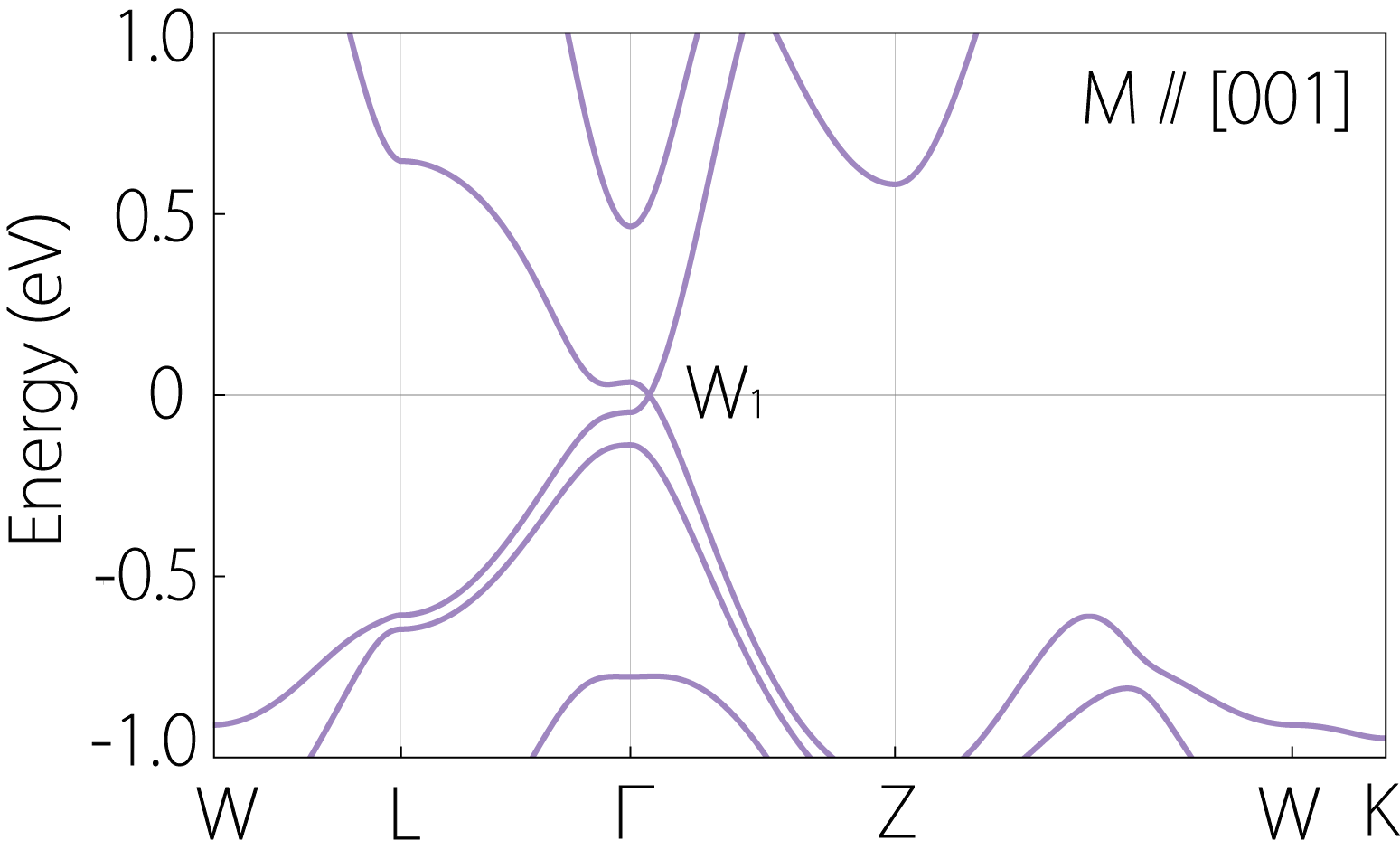}
\par\end{centering}
\caption{\label{Fig_4}Band structure of KCrTe with SOC for magnetization along $[001]$ direction.}
\end{figure}

Next, we study the band structure in the presence of SOC. With SOC included, the band structure will depend on the magnetization direction. Here, we first set the magnetization vector to be along the $[001]$ direction. The results for other directions will be discussed in a while. The calculated band structure for KCrTe is shown in Fig.~\ref{Fig_4}. The overall features of the bands do not change much by SOC. The low-energy bands are still strongly spin polarized along the magnetization direction. One can see that the original quadratic contact point is lifted by SOC. There appears a local gap $\sim$77~meV at $\Gamma$ point. Nevertheless, the system is not fully gapped. There appears a new linear band crossing point $W_1$ on $\Gamma$-Z path. Thus, the system remains to be a semimetal. In the next section, we demonstrate that $W_1$ is a Weyl point and present a detailed analysis of this state.

\begin{figure}
\begin{centering}
\includegraphics[width=8.6cm]{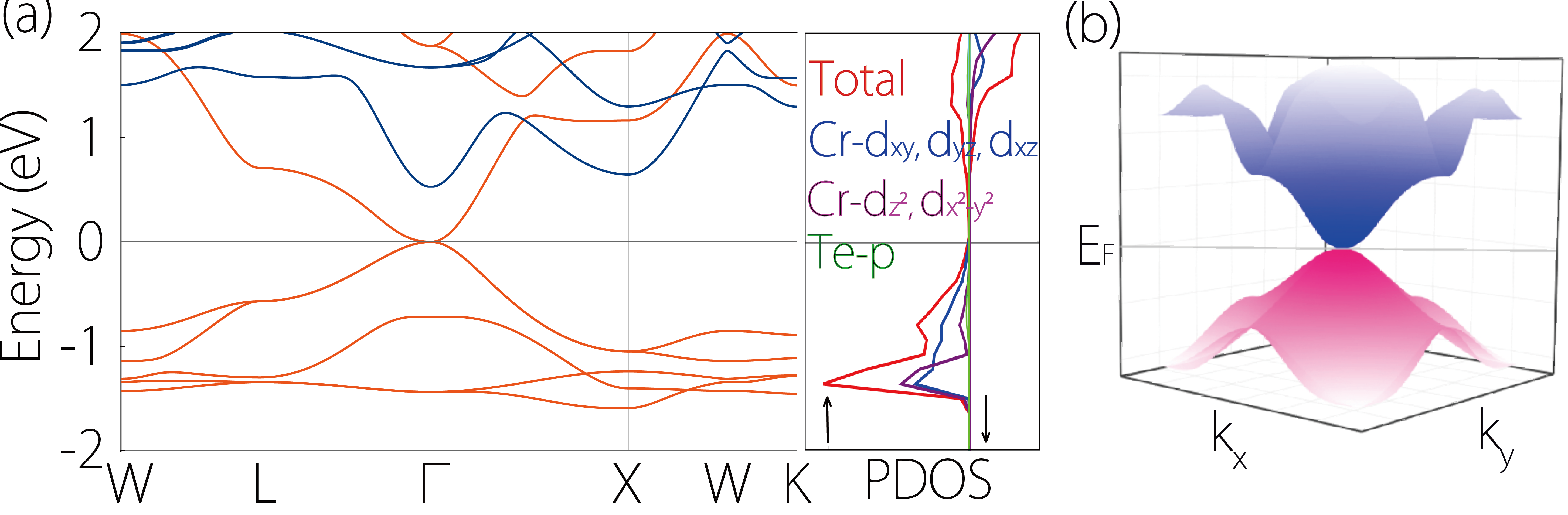}
\par\end{centering}
\caption{\label{Fig_3}(a) Band structure (left) and projected density of states (PDOS, right) of KCrTe without SOC. The orange and blue lines in the band structure indicate the spin-up and spin-down bands, respectively. (b) Band dispersion around the Fermi point. Here, the blue and red colors correspond to conduction and valence bands, respectively.}
\end{figure}

\subsection{Ideal Weyl-half-semimetal phase}
\begin{figure}[b]
\begin{centering}
\includegraphics[width=8.6cm]{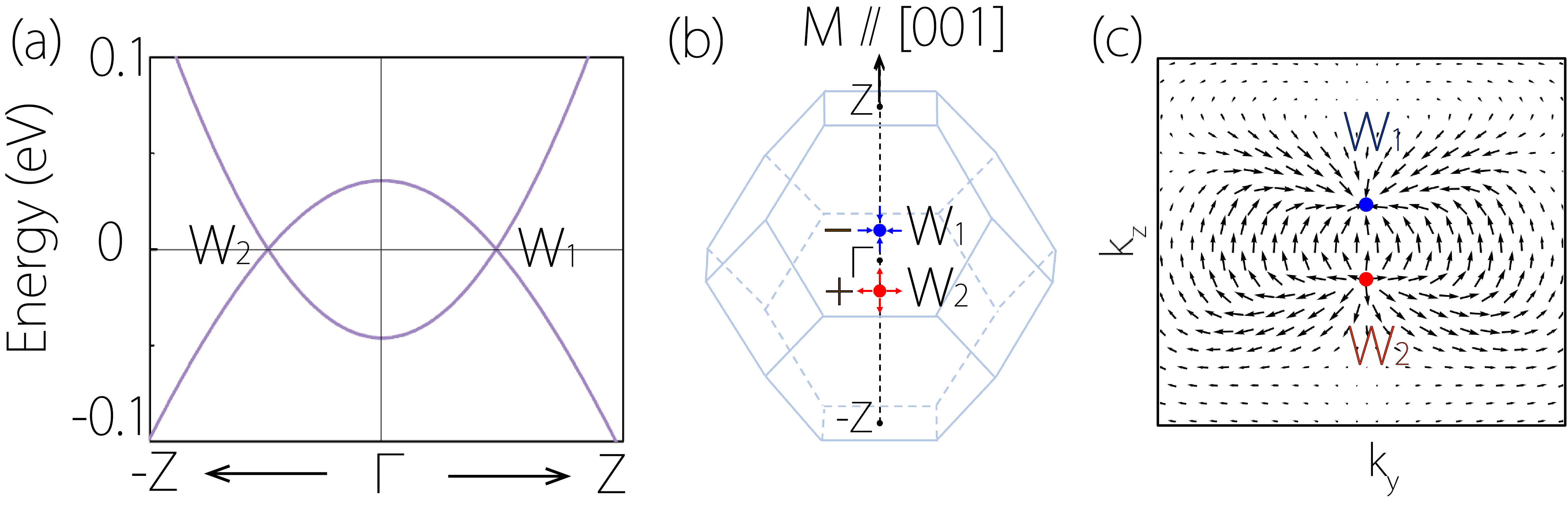}
\par\end{centering}
\caption{\label{Fig_5}(a) Band structure along $\Gamma$-Z path. $W_1$ and $W_2$ are the two Weyl points. (b) Location of the Weyl points in the bulk Brillouin zone for magnetization along [001]. $+$ and $-$ indicate their chirality. (c) Distribution of Berry curvature field in the $k_x=0$ plane.}
\end{figure}

The Fermi point $W_1$ in Fig.~\ref{Fig_5}(a) is formed by the linear crossing between two non-degenerate bands, so it is twofold degenerate. Besides $W_1$, we find there is another crossing point $W_2$ located at $-W_1$ position, which is connected to $W_1$ by $S_{4z}$ symmetry. We have carefully scanned the whole BZ for states around Fermi level, which confirms that $W_1$ and $W_2$ are isolated and they are the only points that make up the Fermi surface. The locations of the two points are given in Table~\ref{tab_ii}. This demonstrates that $W_1$ and $W_2$ are Weyl points, and FM KCrTe is a magnetic Weyl semimetal. Importantly, there is only a single pair of Weyl points in this system, which achieves the minimal possible number of Weyl points. The two Weyl points lie precisely at the Fermi level due to band filling requirement; and the Fermi surface is clean, without presence of any other state. These features make KCrTe an ideal Weyl semimetal material. 

An important characteristic of a Weyl point is its chirality (or Chern number). From DFT results and \textit{ab initio} Wannier model, we determine the Chern number of $W_1$ and $W_2$ to be $-1$ and $+1$, respectively, which fulfills the no-go theorem. The opposite chirality of the two Weyl points can also be visualized from the Berry curvature field around them. In Fig.~\ref{Fig_5}(c), we plot the distribution of Berry curvature field in the $k_x=0$ plane. Here, the Berry curvature is computed by $\boldsymbol{\Omega}({\boldsymbol{k}})=\sum_{n\in occ.}\nabla_{\boldsymbol{k}}\times\left\langle u_{n\boldsymbol{k}}|i\nabla_{\boldsymbol{k}}|u_{n\boldsymbol{k}}\right\rangle$, where $\left|u_{n\boldsymbol{k}}\right\rangle$ is cell-periodic Bloch state and the summation is for all occupied bands below Fermi level. From Fig.~\ref{Fig_5}(c), one can see that the $W_1$($W_2$) acts as a sink (source) of Berry curvature field, which manifests their chirality being $-1$ ($+1$).

\begin{figure}
\begin{centering}
\includegraphics[width=8.6cm]{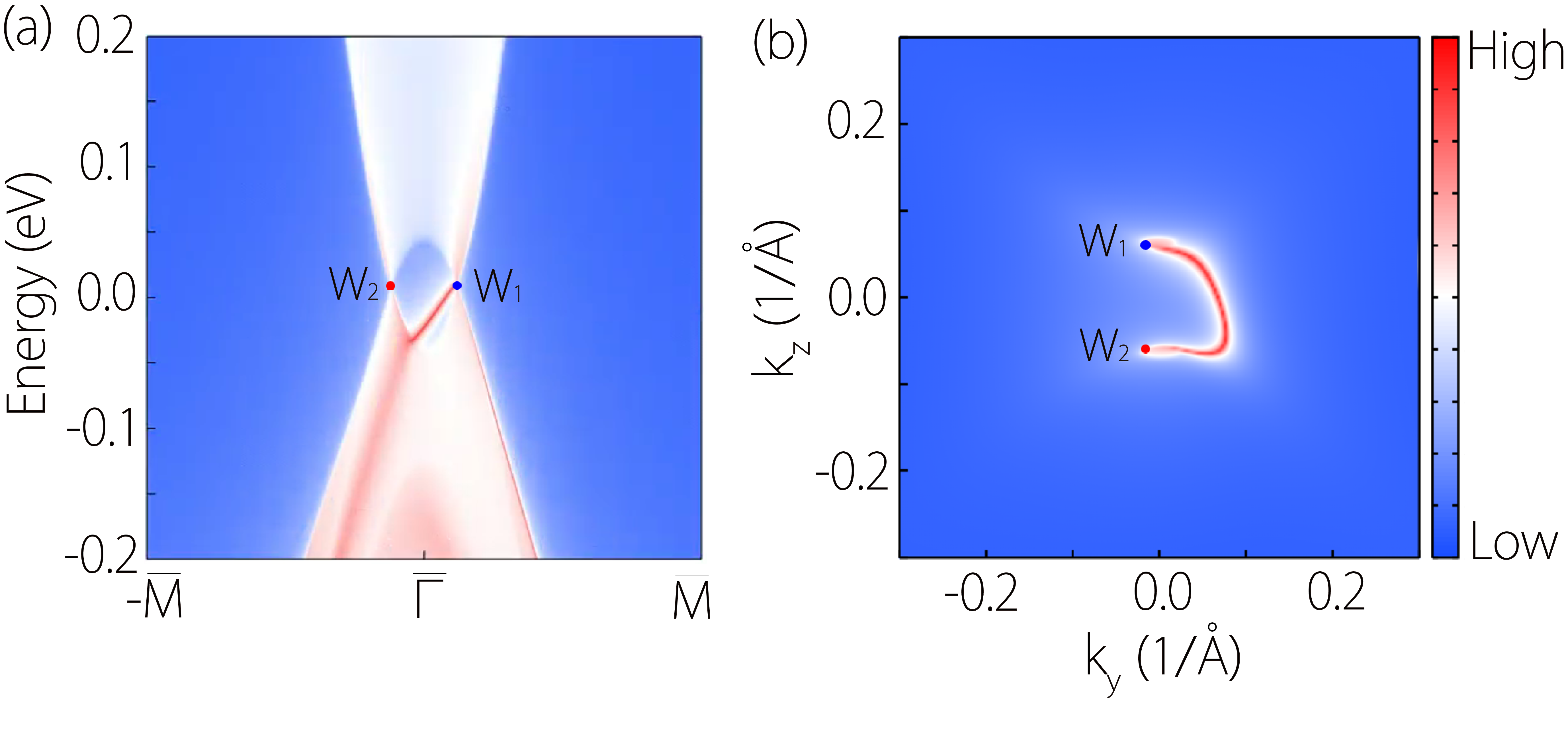}
\par\end{centering}
\caption{\label{Fig_6}(a) Projected surface spectrum for the $(100)$ surface. Surface projections of bulk Weyl points are indicated. (b) Fermi arc on the $(100)$ surface. The dots illustrate the surface projections of the two Weyl points.}
\end{figure}

A hallmark of Weyl semimetal is the topological Fermi-arc surface states~\cite{Chiu2016Classification}. In Fig.~\ref{Fig_6}(a), we plot the calculated surface spectrum on the $(100)$ surface. One can see the existence of a surface band connecting the surface projections of the bulk Weyl points. In Fig.~\ref{Fig_6}(b), we Fermi contour of the surface spectrum, which shows a single Fermi arc connecting the projected Weyl points. We remark that the result in Fig.~\ref{Fig_6}(b) is the simplest Fermi arc pattern for a Weyl semimetal state, which results from the ideal bulk Weyl state. In comparison, in most studied Weyl semimetal materials, the surface Fermi arc patterns are quite complicated, such as those found in TaAs family materials~\cite{Weng2019Magnetic}. Such complication is a direct consequence of (1) multiple pairs of Weyl points in the bulk, (2) Weyl points not located at Fermi level, and (3) the presence of trivial bands at Fermi level. Factor (1) leads to multiple Fermi arcs (which may also form complex connecting pattern). Factors (2) and (3) lead to a background of bulk states (electron or hole pockets). Now, KCrTe (and RbCrTe) is free of these shortcomings, making it an ideal platform to study the fascinating physics predicted for Fermi arcs. 

\subsection{Magnetic control of Weyl points}
In the discussion above, we have shown that the magnetic anisotropy in $X$CrTe ($X$=K, Rb) is relatively weak. This implies that it would be easy to control the orientation of the magnetization vector, e.g., by a small external magnetic field. Rotating the magnetization vector could be a convenient way to tune the band structure, particularly the positions of the Weyl points. 

We have discussed the case with magnetization along $[001]$ direction. One can easily see that if we reverse the magnetization vector, the two Weyl points should switch their chirality by a time reversal operation, namely, $W_1$ would change to chirality $+1$ while $W_2$ change to $-1$. If we rotate the magnetization vector to be along $[100]$ (or $[010]$) direction, the two Weyl points will move to the $\Gamma$-X (or $\Gamma$-Y) path. 

\begin{figure}
\begin{centering}
\includegraphics[width=8.6cm]{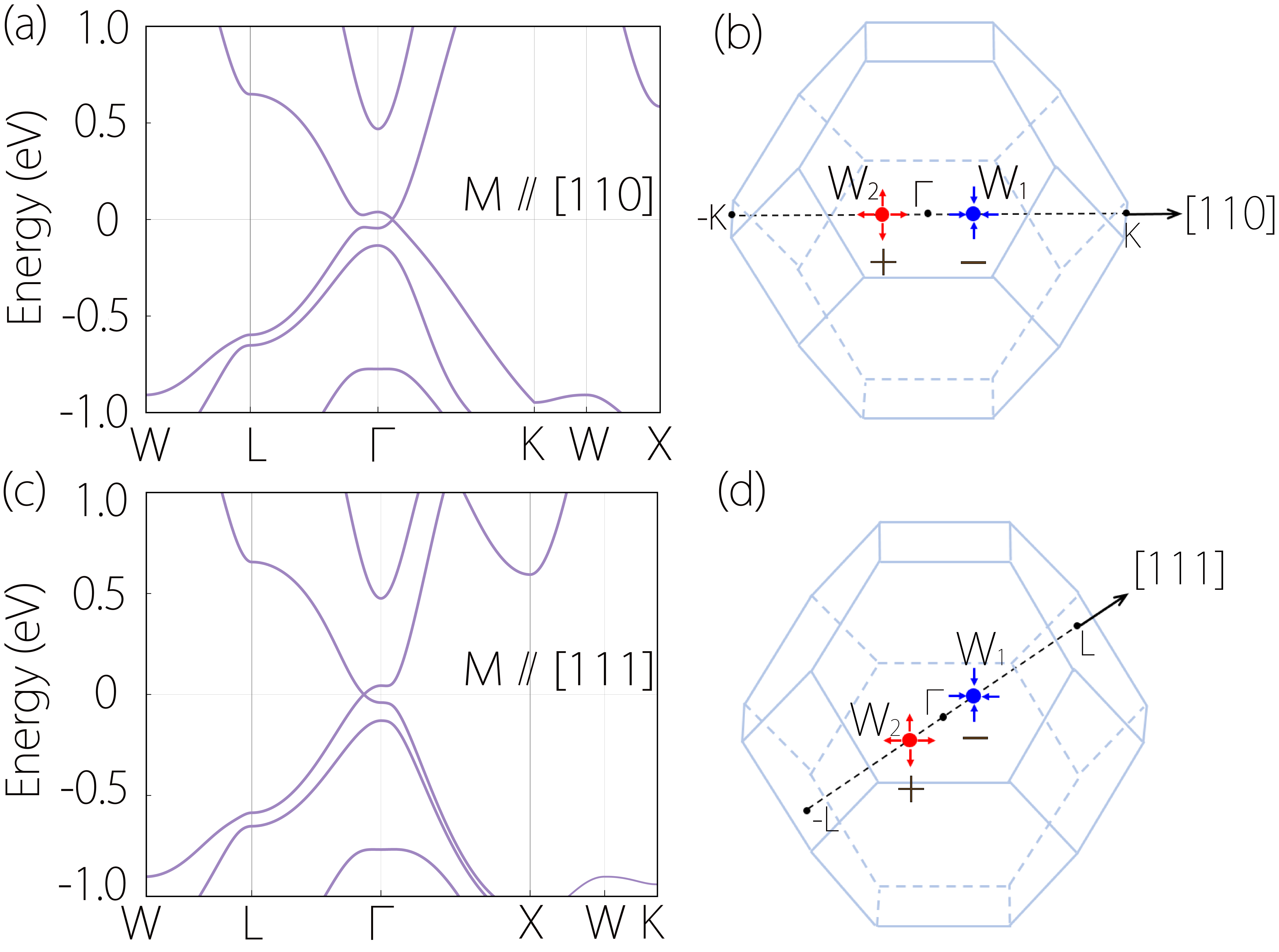}
\par\end{centering}
\caption{\label{Fig_7}Band structures of KCrTe and distribution of the two Weyl points in BZ for magnetization vector along (a)-(b) $[110]$ and (c)-(d) $[111]$ directions.}
\end{figure}

Next, we consider the magnetization vector along the $[110]$ and $[111]$ directions. Interestingly, we find that the single pair of Weyl points are preserved, their energies still pinned at Fermi level, while their positions will move to the high symmetry axis (through $\Gamma$) parallel to the magnetization direction. As shown in Fig.~\ref{Fig_7}(a-d), for $[110]$ case, the two Weyl points are on the $\Gamma$-K path, whereas for $[111]$ case, they are on the $\Gamma$-L path. Their detailed coordinates and chirality are presented in Table~\ref{tab_iii}.

From these results, we see that, first, the ideal Weyl semimetal state is robust under different magnetization directions. This is in contrast top reviously reported magnetization-tunable Weyl semimetals Refs.~\cite{shekhar2018anomalous,kondo2023field}, where the number and{/}or energy of Weyl points would change with magnetization direction. Second, the position of the Weyl points can be flexibly moved by controlling the magnetization direction. This movement will also affect surface Fermi arcs as well as transport properties. Such an interesting effect could be useful for applications in spintronics.

\begin{table}
	\caption{\label{tab_iii}Locations (in units of \AA$^{-1}$) and chiralities of the single pair of Weyl points in KCrTe for configurations with magnetization along $[110]$ and $[111]$ directions.}
	\begin{ruledtabular}
		\begin{tabular}{cccccc}
			Direction & Weyl points & $k_x$ & $k_y$ & $k_z$ & Chirality \\
			\hline 
			$[110]$ & $W_1$ &  ~0.061 &  ~0.061 &  0.0   & $-1$ \\
			& $W_2$ & -0.061 & -0.061 &  0.0   & $+1$ \\
			$[111]$ & $W_1$ &  ~0.059 &  ~0.059 &  ~0.059 & $-1$ \\
			& $W_2$ & -0.059 & -0.059 & -0.059 & $+1$ \\
		\end{tabular}
	\end{ruledtabular}
\end{table}

\subsection{Discussion}
The simple pattern of surface Fermi arcs in Fig.~\ref{Fig_6} also provides us the information of anomalous Hall transport of the system. Denote the positions of the two Weyl points by $\boldsymbol{k}_{W_i} (i=1,2)$. For the case in Fig.~\ref{Fig_6}(b) (where magnetization is along $[001]$ direction), we have $\boldsymbol{k}_{W_1}=(0,0,q)$ and $\boldsymbol{k}_{W_2}=(0,0,-q)$, with $q=0.066$~\AA$^{-1}$. The Fermi arc pattern indicates that for any constant $k_z$ slice of BZ with $|k_z|<q$, namely, when the slice lies between the two Weyl points, it will carry a nonzero Chern number and contribute to the anomalous Hall effect. Indeed, the surface Fermi arc can be regarded as formed by the chiral edge zero modes of all such slices. This offers an easy way to evaluate the anomalous Hall conductivity. As shown in Ref.~\cite{Halperin1987Possible}, the intrinsic anomalous Hall conductivity of a 3D system can be expressed as 
\begin{equation}
\sigma_{ij}=\frac{e^2}{2\pi h} \epsilon_{ijk} v_k,
\end{equation}
where $\epsilon_{ijk}$ is the totally antisymmetric tensor, and $\boldsymbol{v}$ is known as the Chern vector. For an ideal Weyl semimetal, the Chern vector is determined by the location and chirality of Weyl points~\cite{Yang2011Quantum}, with
\begin{equation}
\boldsymbol{v}=\sum_i \chi_i \boldsymbol{k}_{W_i},
\end{equation}
where the summation is over all Weyl points and $\chi_i$ is the chirality of Weyl point $W_i$. From this relation, we can easily find for KCrTe with magnetization along $[001]$ direction, $\sigma_{xy}=-\frac{e^2}{\pi h}q=-81.4$~S/cm. When the magnetization vector is rotated, as we have seen, the Chern vector also changes, leading to the change of Hall response. For example, for the case with magnetization along $[111]$ direction, by using the information in Table~\ref{tab_iii}, we find that the Hall conductivity is changed to $\sigma_{xy}=-72.8~$S/cm. 

In $X$CrTe ($X$=K, Rb), it is important to note that these materials are half metals, and the low energy Weyl fermion states are spin polarized in one spin channel. Although SOC introduces hybridization between spin up and spin down states, the low-energy states are still strongly spin polarization. For example, in KCrTe, we checked that the spin polarization for states around Weyl points remains nearly 100$\%$ under SOC. Such spin-polarized Weyl state is desired for implementing Weyl physics in spintronics applications. 

Finally, besides KCrTe and RbCrTe, we have also investigated several of their sister compounds in the half-Heusler family, including LiCrTe, NaCrTe, KCrS, and KCrSe. We note that LiCrTe, NaCrTe, KCrS, and KCrSe prefer to crystalize in the type-II configuration, which is different from KCrTe and RbCrTe (type-I configuration). The optimized lattice constants of LiCrTe, NaCrTe, KCrS, and KCrSe are 6.57, 6.93, 6.77, and 7.01~\AA, respectively. Their band structures without and with SOC are shown in the Supplementary Materials. The overall features are similar to KCrTe and RbCrTe. Nevertheless, their band structures around Fermi level are not as “clean" as that of KCrTe and RbCrTe.

\section{Conclusions}
In conclusion, we find ideal Weyl-half-semimetal state in the half-Heusler compounds $X$CrTe ($X$=K, Rb). We show these materials have FM ground state, with relative high $T_{\text{C}}$. In the absence of SOC, they are quadratic-contact-point half semimetals, with a single threefold degenerate quadratic nodal point as its Fermi point. Including SOC makes them ideal Weyl half semimetals. They display the following desired features. First, there is only a single pair of Weyl points in these systems, which is the minimal number of Weyl points for a Weyl semimetal. Second, the Weyl points are exactly located at the Fermi level enforced by band filling condition. Third, there is no other trivial bands around Fermi level, generating a “clean” environment for studying the intrinsic Weyl features. Fourth, the Weyl points are strongly spin polarized, making these materials useful for spintronics applications. In addition, we show that the ideal bulk topological features lead to the simplest surface Fermi arc pattern. The position of the Weyl points and the Fermi arc can be readily controlled by rotating the magnetization vector. This also changes the Chern vector and the anomalous Hall response of the system in a simple way. These results will facilitate the study of the fascinating Weyl physics and its application in spintronics. 

\begin{acknowledgments}
The authors thank D. L. Deng and D. Zhao for helpful discussions. This work was supported by National Nature Science Foundations of China (Grant No.~52271185 and No.~51701138), Natural Science Foundation of Tianjin City (Grant No.~17JCQNJC02800), and UM Start-up Grant (No.~251 SRG2023-00057-IAPME).
\end{acknowledgments}

\bibliographystyle{apsrev4-2}
\bibliography{ref}

\end{document}